\documentclass[ejs]{imsart}

\usepackage{hyperref}
\usepackage{amssymb,amsmath,color,bm}
\usepackage{multirow}
\RequirePackage[numbers]{natbib}
\usepackage{url}
\usepackage{booktabs}
\usepackage{subfigure}
\usepackage{graphicx}
\usepackage{enumerate}

\doi{10.1214/18-EJS1421}
\pubyear{2018}
\volume{12}
\firstpage{1150}
\lastpage{1180}

\startlocaldefs\def\y{\mathbf{y}}
\def\Y{\mathbf{Y}}
\def\I{\mathbf{I}}
\def\U{\mathbf{U}}
\def\X{\mathbf{X}}
\def\XX{\mathbb{X}}
\def\E{\mathbf{E}}
\def\EE{\mathbb{E}}
\def\V{\mathbf{V}}
\def\A{\mathbf{A}}
\def\B{\mathbf{B}}
\def\v{\mathbf{v}}
\def\Vmat{\mathbf{V}_{\text{mat}}}
\def\kr{\mbox{kr}}
\def\F{\mathbf{F}}
\def\log{\mbox{log}}
\def\det{\mbox{det}}
\def\tr{\mbox{tr}}
\def\tp#1{[\![#1]\!]}
\def\diag{\mbox{diag}}
\def\W{\mathbf{W}}

\newtheorem{theorem}{Theorem}

\endlocaldefs

\begin{document}

\begin{frontmatter}

\title{Supervised multiway factorization}
\runtitle{Supervised multiway factorization}
\thankstext{T1}{}



\runauthor{Lock and Li}

\author{\fnms{Eric F.} \snm{Lock}\ead[label=e1]{elock@umn.edu}}

\address{Division of Biostatistics, School of Public Health\\ University of Minnesota\\
Minneapolis, MN 55455\\
\printead{e1}\\
}

\author{\fnms{Gen} \snm{Li}\ead[label=e2]{gl2521@cumc.columbia.edu}}

\address{Department of Biostatistics, Mailman School of Public Health \\
 Columbia University\\
New York, NY 10032 \\
\printead{e2}}

\begin{abstract}
We describe a probabilistic PARAFAC/CANDECOMP (CP) factorization for multiway (i.e., tensor) data that incorporates auxiliary covariates, \emph{SupCP}.  SupCP generalizes the supervised singular value decomposition (SupSVD) for vector-valued observations, to allow for observations that have the form of a matrix or higher-order array.  Such data are increasingly encountered in biomedical research and other fields. We use a novel likelihood-based latent variable representation of the CP factorization, in which the latent variables are informed by additional covariates.  We give conditions for identifiability, and develop an EM algorithm for simultaneous estimation of all model parameters. SupCP can be used for dimension reduction,  capturing latent structures that are more accurate and interpretable due to covariate supervision.  Moreover, SupCP specifies a full probability distribution for a multiway data observation with given covariate values, which can be used for predictive modeling.  We conduct comprehensive simulations to evaluate the SupCP algorithm. We apply it to a facial image database with facial descriptors (e.g., smiling / not smiling) as covariates, and to a study of amino acid fluorescence. Software is available at \url{https://github.com/lockEF/SupCP}.  \end{abstract}


\begin{keyword}
\kwd{Faces in the wild}
\kwd{Dimension reduction}
\kwd{Latent variables}
\kwd{Parafac/Candecomp}
\kwd{Singular value decomposition}
\kwd{Tensors}
\end{keyword}


\tableofcontents

\end{frontmatter}

\section{Introduction}
\label{sec:intro}

Data are often best represented as a \emph{multiway} array, also called a \emph{tensor}, which  extends the familiar two-way data matrix (e.g., Samples $\times$ Variables).  Multiway data are increasingly encountered in many application areas.  For example, in biomedical applications molecular profiling data may be measured over multiple body regions, tissue types, or time points for the same sample set (e.g., Samples $\times$ Molecules $\times$ Time $\times$ Tissue).  A multiway representation is also useful for imaging modalities.  In this article we describe an application to aligned color images of the faces of different individuals, from the Labeled Faces in the Wild database \citep{learned2016labeled}.  In particular we consider images for $4000$ individuals, each with pixel dimensions $90 \times 90$, where each pixel contains three color intensities (red, green, blue); thus the data is given by a 4-way array of dimensions $4000 \times 90 \times 90 \times 3$ (Individual $\times$ $X$ $\times$ $Y$ $\times$ Color).

Unsupervised factorization methods for multiway data, which extend well-known methods such as the \emph{singular value decomposition} (\emph{SVD}) and \emph{principal component analysis} (\emph{PCA}) for a data matrix, are commonly used in many fields.   See \cite{kolda2009tensor} for an accessible and detailed survey of multiway factorization methods and their uses. A classical and straightforward method is the \emph{PARAFAC/CANDECOMP} (\emph{CP}) \citep{harshman1970foundations} decomposition, in which the data are approximated as a linear combination of rank-1 tensors.  Alternative approaches include the \emph{Tucker}  decomposition \citep{tucker1966some}, and the \emph{population value decomposition} (\emph{PVD}) for 3-way data \citep{crainiceanu2012population}.  All of these approaches reconstruct the data using a small number of underlying patterns in each modality, and they are therefore useful for dimension reduction, de-noising, and exploratory analysis.

In many scenarios, additional variables of interest are available for multiway data objects.  For example, \cite{kumar2009attribute} provide several attributes for the images in the Faces in the Wild database, which describe the individual (e.g., gender and race) or their expression (e.g., smiling / not smiling).  There is a growing literature on analyzing the relationship between multiway data and additional variables via predictive models in which the multiway data are predictors and the additional variables are treated as outcomes.  \cite{zhou2013tensor} propose a tensor regression model for a clinical outcome, in which the coefficients for multiway data are assumed to have the structure of a low-rank CP decomposition.  \cite{miranda2015} describe a Bayesian formulation for regression models with multiple outcome variables and multiway predictors, which is applied to a neuroimaging study.  Multiple methods have been developed for the classification of multiway data \citep{bauckhage2007robust,tao2007supervised,lyu2016discriminating}, extending well-known linear classification techniques under the assumption that model coefficients have a factorized structure.

As in matrix factorization techniques such as PCA, the goal of multiway factorization is to capture underlying patterns that explain variation in the data.  It is often reasonable to assume that these patterns are at least partially driven by additional measured variables of interest (e.g., gender or expression for facial images).  Therefore, supervising on these variables can improve both the accuracy and the interpretation of the results.  In this article we introduce a probabilistic CP factorization for multiway data that incorporates additional variables, called \emph{SupCP}. To our knowledge there is no previous work wherein additional variables inform the factorization or dimension reduction of multiway data.  Moreover, SupCP specifies a full probability distribution for a multiway data observation with given covariate values.  It can therefore be used for modeling of a multiway outcome given vector-valued predictors (rather than predicting an outcome vector from multiway predictors), for which there is little existing methodology.

  There is an existing literature on supervised factorization methods for two-way data (matrices).  For example, \cite{bair2006prediction} describe an approach in which PCA is performed on a subset of variables that are selected by the strength of their association with an outcome, and the resulting principal components are used in a predictive model for the outcome. However, in our framework the additional variables are not outcomes to be predicted from multiway data; rather, they are treated as auxiliary covariates to inform a CP model for multiway data. In this respect our framework is most related to the \emph{supervised SVD} (\emph{SupSVD}) approach \citep{li2016supervised} for matrices, which has also been generalized to accommodate sparse or functional PCA models \citep{li2015supervised} and non-parametric relations between the covariates and principal components \citep{fan2016projected}.

The rest of this manuscript is organized as follows.  In Section~\ref{notation} we introduce some algebraic notation and terminology for multiway factorization.  In Section~\ref{sec:mod} we introduce the SupCP model, in Section~\ref{likelihood} we give its full likelihood, and in Section~\ref{ident} we describe its identifiability properties.  In Section~\ref{est} we describe a modified EM-algorithm for simultaneous estimation of all parameters in the model. In Section~\ref{othermeth} we describe the connections between SupCP and other methods in more detail.  In Section~\ref{sims} we describe a comprehensive simulation study to evaluate the SupCP method, in Section~\ref{appAmino} we apply it to a study of amino acid fluorescence, and in Section~\ref{app} we describe an application to facial image data from the Faces in the Wild database.

\section{Notation and Preliminaries}
\label{notation}

Here we give some algebraic notation and terminology for multiway data that will suffice to describe the development in the following sections.  For a general and more comprehensive introduction to multiway data and its factorization we recommend \cite{kolda2009tensor}.

Define a $K$-way array (i.e., a $K$th-order tensor) by $\XX: d_1 \times \hdots d_K$, where $d_k$ is the dimension of the $k$th \emph{mode}.  The entries of the array are given by square brackets, $\XX[i_1,\hdots,i_K]$, where $i_k \in \{1,\hdots,d_k\}$ for $k \in 1,\hdots, K$.  In this article we use script capital letters to denote arrays of order greater than $2$, bold capital letters to denote matrices ($\X: d_1 \times d_2$), and bold lowercase letters to denote vectors ($\v: d \times 1$).

For vectors $\v_1,\hdots,\v_K$ of length $d_1,\hdots, d_K$, respectively, define the \emph{outer product}
\[
\XX = \v_1 \circ \v_2 \cdots \circ \v_K
\]
as the $K$-way array of dimensions $d_1 \times \hdots \times d_K$, with entries
\[\XX[i_1,\hdots,i_K] = \prod_{k=1}^K \v_{k}[i_k]. \]
An array given by the outer product of vectors has \emph{rank} 1.
For matrices $\V_1, \hdots, \V_K$ of the same column dimension $R$, we introduce the notation
\[ \tp{\V_1,\hdots,\V_K} = \sum_{r=1}^R \v_{1r} \circ \cdots \circ \v_{Kr},\]
where $\v_{kr}$ is the $r$th column of $\V_k$.  This defines a rank $R$ CP factorization.  Often, the columns of each $\V_k$ are scaled to have unit norm ($\|\v_{kr}\|_F=1$ for $r=1,\cdots,R$, where $\|\cdot\|_F$ represents the Frobenius norm) and scalar weights $\lambda_r$ are given for each rank-$1$ component:
\[\sum_{r=1}^R \lambda_r \cdot \v_{1r} \circ \cdots \circ \v_{Kr}.\]
The rank $R$ CP approximation to an array of higher-rank $\XX$ is typically estimated by minimizing the residual sum of squares
\begin{align}
\|\XX-\sum_{r=1}^R \lambda_r \cdot \v_{1r} \circ \cdots \circ \v_{Kr}\|_F^2.	 \label{eqParafac}
\end{align}
From (\ref{eqParafac}) it is easy to see how the CP factorization extends the SVD for arrays of order greater than $2$.  However, there are some important differences.  In particular, for $K>2$ it is not common to enforce orthogonality among the columns of $\V_k$ because this restricts the solution space and is not necessary for identifiability.  Also, while the dimensions of the row space and column space are equal for $K=2$, there is no analogous symmetry for $K>2$, and therefore the rank of $\XX$ may be greater than the dimension of each mode (i.e., $d_1,\hdots,d_K$).

Finally, it is often useful to represent a multiway array in matrix form, wherein it is \emph{unfolded} along a given mode.  For this purpose we let $\X^{(k)}: d_k \times \left(\prod_{i \neq k} d_i \right)$ be the array $\XX$ arranged so that the $j$th row contains all values for $j$th \emph{subarray} in the $k$th mode;  specifically, for $k=1$
\[\X^{(1)}\left[j,i_2+\sum_{k=3}^K \left(\prod_{l=2}^{k-1}d_l\right) (i_k-1)\right] = \XX[j,i_2,\hdots,i_K],\]
and we define $\X^{(k)}$ similarly for $k=2,\hdots,K$.

\section{Model}
\subsection{Latent Variable Model}
\label{sec:mod}

Suppose we observe a $K$-way data array (e.g., an image with modes $x$ $\times$ $y$ $\times$ color) for each of $n$ observational units, hereafter called \emph{samples}.  The full data set can then be represented as a $(K+1)$-way array with samples as the first mode, $\XX: n \times d_1 \times \cdots \times d_K$. In addition, we observe a $q$-dimensional vector of covariates for each sample (e.g., describable features for facial images), given by the matrix $\Y: n \times q$. Without loss of generality, we assume both $\XX$ and $\Y$ are centered across samples.

The SupCP model extends the CP factorization for $\XX$ \eqref{eqParafac} to a generative likelihood model in which the covariates in $\Y$ inform latent patterns for the sample mode.  Specifically, the rank $R$ SupCP model is
\begin{align}
\label{eqSupCP}
\begin{split}
\XX &= \tp{\U,\V_1, \hdots,\V_K} + \EE  \\
\U &= \Y \mathbf{B} +\F,
\end{split}
\end{align}
where
\begin{itemize}
\item $\U: n \times R$ is a latent score matrix for the samples,
\item $\{\V_k: d_k \times R\}_{k=1}^K$ are loading matrices in different modes,
\item $\EE: n \times d_1 \times \cdots \times d_K$ is an error array with independent normal entries $N(0,\sigma_e^2)$,
\item $\mathbf{B}: q \times R$ is a regression coefficient matrix for $\Y$ on $\U$,
\item $\F: n \times R$ has independent and identically distributed (iid) multivariate normal rows with mean $\mathbf{0}$ and $R \times R$ covariance $\Sigma_f$.
\end{itemize}

The loadings $\tp{\V_1, \hdots,\V_K}$ form a basis for the sampled $d_1 \times \cdots \times d_K$ arrays, that can be decomposed into patterns for each mode.  The scores $\U$  reconstruct each sample from this basis, and give an efficient lower-dimensional embedding of $\XX$ if the data have multilinear structure.  Thus, to describe the relation of covariates $\Y$ on structured variation in $\XX$ it suffices to consider the relation between $\Y$ and $\U$, and this gives an efficient generative model for $\XX$ given $\Y$.  Moreover, the patterns in each mode (${\V_1,\hdots,\V_K}$) are often interpretable, and supervision on $\Y$ can facilitate this interpretation by capturing how they relate to the covariates $\Y$ (see, e.g., Section~\ref{appAmino}).

The sample scores $\U$ are decomposed into what is explained by the covariates ($\Y \mathbf{B}$) and not explained by the covariates ($\F$).  Therefore, the SupCP model decomposes the variation in $\XX$ into variation explained by the covariates, structured variation not explained by the covariates, and independent residual noise;  this is easily seen with the following reformulation of \eqref{eqSupCP}:
\[\XX = \tp{\mathbf{Y B}, \V_1, \hdots, \V_K} \, + \, \tp{\F, \V_1 , \cdots, \V_K} \, + \, \EE.\]
The model allows for components that have little relation to $\Y$, are partially explained by $\Y$, or that are almost entirely explained by $\Y$, depending on the relative size of $\B$ and $\Sigma_f$.  If $\XX$ has structured low-rank variation that is orthogonal to $\Y$, then for some components $r$ the sample variance of the $r$th column of $\Y\B$ will be very small relative to $\Sigma_f[r,r]$.  Also, if $\|\Sigma_f\|_F$ goes to $0$, it implies that all structured variation in $\XX$ is dependent on the variables in $\Y$.  In Section~\ref{est} we describe a maximum likelihood estimation approach for all parameters simultaneously, including $\Sigma_f$ and  $\sigma_e^2$; therefore, the appropriate level of covariate supervision is determined empirically.

In practice, we constrain the columns of each $\V_k$ to have a unit norm. The scaling of the factorization components, given by the $\lambda_r$'s in \eqref{eqParafac}, are therefore subsumed into $\U$.  Remarkably, this is enough to make the model parameters identifiable up to trivial sign changes and ordering of components, as discussed in Section~\ref{ident}.

\subsection{Likelihood}
\label{likelihood}

The full marginal distribution for $\XX$ under the SupCP model, specified in \eqref{eqSupCP}, is a multilinear normal distribution \citep{ohlson2013multilinear}.  Here we derive its equivalent likelihood by considering the more tractable matricized version of $\XX$, $\X^{(1)}$.  The $i$th row of $\X^{(1)}$ gives all values for the $i$th sample.  We define $d$ as the overall dimension for each sample, \[d = \prod_{k=1}^K d_k,\]
so $\X^{(1)}$ is of dimension $n \times d$.  Further, let $\Vmat: d \times R$ be a matricized version of $\V_1,\hdots,\V_K$, where the $r$th column is the vectorization of the $K$-way array given by $\v_{1r} \circ \cdots \circ \v_{Kr}$.
This gives a matrix representation of the SupCP model,
\begin{align}
\X^{(1)} = \Y \B \Vmat^T + \F \Vmat^T + \E^{(1)}. \label{eqSupCPmat}
\end{align}
It follows that the distribution of $\X^{(1)}$, marginalizing over $\F$, is multivariate normal with log-likelihood
\begin{align} L(\X^{(1)}) \propto -\frac{n}{2} \log\, \det (\Sigma_\X) -\frac{1}{2} \tr\left((\X^{(1)}-\mu_\X)\Sigma_\X^{-1} (\X^{(1)}-\mu_\X)^T\right)	\label{eqLike}
\end{align}
where
\[\Sigma_\X = \Vmat \Sigma_f \Vmat^T+\sigma^2_e \I_d,\]
and
\[\mu_\X = \Y \B \Vmat^T,\]
and $\I_d$ is a $d\times d$ identity matrix.
Because the likelihoods of $\XX$ and $\X^{(1)}$ are equivalent, it suffices to consider \eqref{eqLike}.  However, this likelihood function is very high-dimensional and not convex with respect to all parameters. Thus it is difficult to optimize directly over all parameters.  In Section~\ref{est} we describe an {\em Expectation-Maximization} (EM) algorithm to maximize this likelihood over $\{\V_1,\hdots,\V_K, \Sigma_f,$ $\sigma^2_e, \B\}$.

The covariance matrix for the $d-dimensional$ matricized data for each sample is given by $\Sigma_\X$.  This can be described as a spiked covariance model with rank$-R$ structure and independent noise.  The resulting covariance is not in general \emph{separable} \cite{hoff2011separable}, meaning that it can not be factorized into the Kronecker product of separate covariances for each mode.  Indeed, our model assumes that structured variation in $\XX$ is driven by latent factors that affect multiple modes, rather than an independent factor structure for each mode \cite{fosdick2014separable}.

\subsection{Identifiability}
\label{ident}

Here we describe conditions under which the parameters in the SupCP model are identifiable.  That is, for two sets of parameters
\[\Theta = \{\V_1,\hdots,\V_K, \Sigma_f, \sigma^2_e, \B\} \, \text{ and } \,  \tilde{\Theta} = \{\tilde{\V}_1,\hdots,\tilde{\V}_K, \tilde{\Sigma}_f, \tilde{\sigma}^2_e, \tilde{\B}\},\] we give conditions for which equivalency of the likelihoods, $L(\XX \mid \Theta)=L(\XX \mid \tilde{\Theta})$, implies equivalency of the parameters, $\Theta = \tilde{\Theta}$.

Clearly, the model is not identifiable under arbitrary scaling of $\B$ and the $\V_k$'s.  For example, if $\{\Sigma_f, \sigma^2_e, \B\} =\{\tilde{\Sigma}_f, \tilde{\sigma}^2_e, \tilde{\B}\}$ and $\tilde{\V}_k = a_k \V_k$ for $k=1,\hdots,K$ where $\prod_{k=1}^K a_k =1$, then $L(\XX \mid \Theta)=L(\XX \mid \tilde{\Theta})$.  We address this by scaling the columns in $\V_k$ to have unit norm, and restricting the first nonzero entry (typically the first entry) of each column to be positive:
\begin{enumerate}[(a)]
\item $\|\v_{kr}\|_F = 1$, and $\V_k[1,r] > 0$ for $k=1,\hdots,K$ and $r=1,\hdots,R$.	
\end{enumerate}
The model is also clearly not identifiable under permutation of the $R$ rank-1 components, i.e., applying the same re-ordering to the columns in $\V_1,\hdots,\V_K$, and $\B$.  This is easily addressed by rank-ordering the components under any given criteria;  by default we order the components according to the diagonal entries of $\Sigma_f$, which will be distinct for non-trivial cases:
 \begin{enumerate}[(b)]
\item $\Sigma_f[r_1,r_1]>\Sigma_f[r_2,r_2]$ for $r_1>r_2$.
\end{enumerate}
Alternatively the components can be ordered by their variance explained by $\Y$, $(\Y \B)^T (\Y \B) [r,r]$, or by their overall variance, $(\Y \B)^T (\Y \B) [r,r] + \Sigma_f[r,r]$.

After scaling and ordering, the model is identifiable under general conditions for $K \geq 2$.  It is remarkable that no other restrictions are required.  In particular, the columns of $\V_k$ need not be orthogonal and  $\Sigma_f$ need not be diagonal for identifiability.

There is a vast body of literature on the uniqueness of components in a CP factorization up to scaling and ordering (see Section 3.2 of \cite{kolda2009tensor} for a review), which can be used to derive conditions for identifiability of the model given (a) and (b) above.  Here we use a result of \cite{sidiropoulos2000uniqueness}, which requires the notion of \emph{k-rank}. The $k$-rank of a matrix $\A$, $\kr(\A)$, is defined as the maximum number $k$ such that any $k$ columns of $\A$ are linearly independent.  Note that in general $\kr(\A) \leq \mbox{rank}(\A)$ and if $\A$ is of full column rank, then $\kr(\A) = \mbox{rank}(\A)$.  Sufficient conditions for identifiability of the SupCP model are given in Theorem~\ref{thm1}.

\begin{theorem}
\label{thm1}
For the model defined in~(\ref{eqSupCP}), if
\begin{align}
\kr({\Y\B}) + \sum_{k=1}^K \kr(\V_k) \geq 2R+K, \label{krCond}
\end{align}
and if $\Y$ is of full column rank, then the parameters $\{\V_1,\hdots,\V_K, \Sigma_f, \sigma^2_e, \B\}$ are identifiable under the restrictions (a) and (b).
\end{theorem}

The proof of Theorem~\ref{thm1} is given in Appendix~\ref{app1}.   Generally we expect the solution to be matrices of full rank satisfying $\kr({\Y\B}) = \mbox{min}(n,q,R)$ and $\kr({\V_k}) = \mbox{min}(d_k,R)$, and therefore the model will be identifiable if $R$ is sufficiently small relative to the dimensions of $\XX$.  Nevertheless, the conditions of Theorem~\ref{thm1} are easily verifiable.  Moreover, these conditions are sufficient but not necessary, and so failure to satisfy~(\ref{krCond}) does not imply that the model is not identifiable.

In practice, although not necessary for identifiability, we also constrain the covariance $\Sigma_f$ to be diagonal:
 \[\Sigma_f[i,j]=0 \, \text{ for }\, i \neq j.\]
This simplifies the model complexity considerably, and tends to improve performance of the estimation algorithm.  We impose this constraint by default, but allow the option of an unconstrained covariance in our implementation.

\section{Model Estimation}
\label{est}

Here we describe estimation of all parameters for the model specified in Section~\ref{sec:mod}.  In Section~\ref{EMalg} we describe a modified EM-algorithm for optimizing the likelihood~(\ref{eqLike}) for fixed rank $R$.  In Section~\ref{rankSelect} we discuss selection of the rank.

\subsection{EM algorithm}
\label{EMalg}

For our implementation of the modified EM algorithm, the latent random variables $\U$ are updated via their conditional moments at each iteration. For this purpose, it is useful to consider the joint log-likelihood of $\XX$ and $\U$:
\begin{align}
 L(\XX,\U) &= L(\XX \mid \U) + L(\U), \label{eqJointLike}
\end{align}
where
\begin{align}
 L(\XX \mid \U)= -n d \, \log \, \sigma^2_e + \sigma_e^{-2} \|\XX- \tp{\U,\V_1,\hdots,\V_k}\|_F^2, \label{eqCondLike}	
\end{align}
and
\begin{align*}
 L(\U)= - n \, \log \,\det \, \Sigma_f -\frac{1}{2} \tr \left((\U-\Y \B) \Sigma_f^{-1} (\U-\Y \B)^T\right). 	
\end{align*}
Moreover, the conditional log-likelihood (\ref{eqCondLike}) can be expressed in terms of the matrices $\X^{(1)}$ and $\Vmat$ used in Section~\ref{likelihood}:
\[ L(\XX \mid \U)= -n d \, \log \, \sigma^2_e + \sigma_e^{-2} \tr \left((\X^{(1)}- \U \Vmat^T)(\X^{(1)}- \U \Vmat^T)^T \right),\]
and so the joint likelihood (\ref{eqCondLike}) gives the sum of two multivariate normal likelihoods.  This facilities a modified EM algorithm, which is described in detail in steps~\ref{init} through~\ref{Mstep}.

\subsubsection{Initialization}
\label{init}
One could initialize $\U$ and $\V_1,\hdots,\V_K$ via a least-squares CP factorization (\ref{eqParafac}), wherein the columns of each $\V_k$ are scaled to have unit norms and the corresponding weights are absorbed into $\U$.  However, this is estimated via an alternating least squares algorithm that can be computationally intensive, and is not guaranteed to find the global minimizer.

Alternatively we consider a random initialization in which the entries of each $\V_k$ are generated independently from a normal distribution with mean $0$, and then the columns of each $\V_k$ are scaled to have unit norm, for $k=1,\hdots,K$.  The latent factor $\U$ is initialized by
\[\U = \X^{(1)} \Vmat.\]
In practice we find that random initialization performs as well as initialization via a CP factorization (see Appendix~\ref{app2}), and so in our implementation we initialize randomly by default, with the option to initialize via CP factorization.

After initializing $\U$ and $\V_1,\hdots,\V_K$, we initialize  $\B$ by regressing $\U$ on $\Y$:
\[\B = (\Y^T \Y)^{-1} \Y^T \U.\]
We initialize $\sigma_e^2$ by the sample variance of the entries in the residual array
$\XX - \tp{\U,\V_1,\hdots,\V_K}$,
and initialize $\Sigma_f$ by the diagonal entries of the covariance of the residuals for the random factors $\F = \U - \Y \B$:
\[\Sigma_f =  \diag \left(\F^T \F/n\right),\]
where $\diag(\A)$ sets the off-diagonal values of a square matrix $\A$ to zero.

\subsubsection{E-Step}
\label{Estep}
We estimate $\U$ via its conditional expectation,  given the data $\{\X^{(1)},\Y\}$ and fixed parameters:
\begin{align}
\widehat{\U} = (\sigma^2_e \Y \B \Sigma_f^{-1} + \X^{(1)^T} \Vmat) (\Vmat^T\Vmat+\sigma^2_e \Sigma_f^{-1})^{-1}. \label{Uhat}
\end{align}
Similarly, the conditional variance of $\U$ is
\[\widehat{\Sigma}_U = \left(\frac{1}{\sigma_e^2} \Vmat^T \Vmat+\Sigma_f^{-1} \right)^{-1}.\]
This matrix inverse can be efficiently evaluated via the Woodbury matrix identity \citep{press1986numerical}. Thus, the computation of the E-step is very  efficient.
\subsubsection{M-Step}
\label{Mstep}

The parameters $\V_1,\hdots,\V_K, \B, \Sigma_f, \sigma^2_e$  are updated by maximizing the conditional expectation of the log-likelihood.  Derivations for the following conditional maximum likelihood estimates, using the expected moments of $\U$ from~\ref{Estep}, are given in Appendix~\ref{app3}.

The updating step for $\V_1$ is
\[\V_1 = \X^{(2)^T} \widehat{\W}_{\text{mat}}^{(1)} \left(\Vmat^{(1)^T} \Vmat^{(1)} \cdot (\widehat{\U}^T \widehat{\U} + n\widehat{\Sigma}_{U}) \right)^{-1},\]
where the $r$th column of $\widehat{\W}_{\text{mat}}^{(1)}: n \prod_{k=2}^K d_k \times R$ is the vectorization of $\widehat{\mathbf{u}}_{r} \circ \v_{2r} \circ \cdots \circ \v_{Kr}$, the $r$th column  of $\Vmat^{(1)}: \prod_{k=2}^K d_k \times R$  is the vectorization of $\v_{2r} \circ \cdots \circ \v_{Kr}$, and $\cdot$ is the Hadamard (entry-wise) product.  The remaining loading matrices $\V_2, \hdots, \V_K$ are updated similarly.

We then update the  parameters $\B$, $\Sigma_f$, and $\sigma^2_e$ by setting their respective partial derivative of the expected log-likelihood equal to $0$. As a result, we obtain:
\begin{align}
\B &= (\Y^T \Y)^{-1} \Y^T \widehat{\U}, \nonumber \\
 \Sigma_f &= \frac{1}{n} \left(\widehat{\U}^T \widehat{\U} + n\widehat{\Sigma}_{U} + (\Y \B)^T(\Y \B)- (\Y \B)^T\widehat{\U}- \widehat{\U}^T(\Y \B) \right), \label{paramUpdates} \\
 \sigma^2_e &= \frac{1}{nd} (\tr(\X^{(1)}(\X^{(1)^T}-2\Vmat\widehat{\U}^T))
	+ n (\Vmat^T\Vmat\widehat{\Sigma}_U)
	+ \tr(\widehat{\U}\Vmat^T\Vmat\widehat{\U}^T) ). \nonumber
\end{align}

As a final step, we adjust the sign and scale of each column of $\V_k$ and re-scale $\B$ and $\Sigma_f$ accordingly.  This step is only for identifiability purpose (see Section~\ref{ident}) and does not change the likelihood.  Specifically, if $\tilde{\V}_1,\hdots,\tilde{\V}_K,\tilde{\V}_{\text{mat}}, \tilde{\B}$ and $\tilde{\Sigma}_f$ correspond to their respective unscaled parameters, the scaled versions are given by
\begin{align*}
	\v_{kr} &=\tilde{\v}_{kr}/\|\tilde{\v}_{kr}\|_F &\text{for } k=1,\hdots,K\text{ and } \, r=1,\hdots,R, \\
	\B[i,r] &= \tilde{\B}[i,r] \cdot \|\tilde{\v}_{\text{mat}_{r}}\|_F &\text{for } i=1,\hdots,q \text{ and } r=1,\hdots,R,    \\ 	
	\Sigma_f[r,s] &= \tilde{\Sigma}_f[s,r] \cdot \|\tilde{\v}_{\text{mat}_{s}}\|_F \cdot \|\tilde{\v}_{\text{mat}_{r}}\|_F &\text{for } s=1,\hdots,R \text{ and } r=1,\hdots,R.
\end{align*}
Optionally, we also restrict $\Sigma_f$ to be diagonal by setting $\Sigma_f[r,s]=0$ for $r \neq s$.
\\

We iterate between the E-step and the M-step of the algorithm until the marginal likelihood (\ref{eqLike}) converges.  The M-step (Section~\ref{Mstep}) maximizes the expected log-likelihood coordinate-wise over each parameter, but may not give a global optimum over the full parameter space $\{\V_1,\hdots,\V_K, \Sigma_f, \sigma^2_e, \B\}$. Nevertheless, the marginal likelihood is guaranteed to increase over each iteration and therefore converge.  In practice the likelihood may converge to a local, rather than a global, optimum, depending on the initial values.  We find that a form of annealing, in which random variation is incorporated into the first several iterations, helps to avoid dependence on the initialization. See Appendix~\ref{app2} for more detail.

After estimation, $\tp{\widehat{\U},\V_1,\hdots,\V_K}$ is the full low-rank reconstruction of $\XX$, where $\widehat{U}$ is given in (\ref{Uhat}). The reconstruction of $\XX$ given $\Y$ only is $\tp{\Y\B,\V_1,\hdots,\V_K}$.

\subsection{Rank selection}
\label{rankSelect}

We recommend selecting the rank $R$ of the model by likelihood cross-validation, in which model parameters are estimated via a training set and evaluated using the likelihood of a test set.  Such an approach is desired because it directly assesses the fit of the specified probabilistic model (via the likelihood) in a way that is robust to overfitting.  Also for exploratory purposes it provides validation that the estimated components describe real underlying structure.

First, we select $n_\text{train}$ samples to fit the model, yielding the reduced data array $\XX_\text{train}: n_\text{train} \times d_1 \times \cdots \times d_K $  and covariate matrix  $\Y_\text{train}: n_\text{train} \times q$.  The remaining samples are considered the test set, yielding $\XX_\text{test}: (n-n_\text{train}) \times d_1 \times \cdots \times d_K $  and  $\Y_\text{test}: (n-n_\text{train}) \times q$.  For candidate values of $R$, we optimize the likelihood of the training data $\XX_\text{train}$ and $\Y_\text{train}$ using the algorithm in Section~\ref{EMalg} to obtain estimates $\{\widehat{\V}_{1,R},\hdots,\widehat{\V}_{K,R}, \widehat{\Sigma}_{f,R}, \widehat{\sigma}^2_{e,R}, \widehat{\B}_{R}\}$.  The resulting estimates are assessed via the log-likelihood of the test set
\begin{align*}
L \left(\XX_\text{test},\Y_\text{test} \mid \widehat{\V}_{1,R},\hdots,\widehat{\V}_{K,R}, \widehat{\Sigma}_{f,R}, \widehat{\sigma}^2_{e,R}, \widehat{\B}_{R} \right)
\end{align*}
which can be evaluated as in \eqref{eqLike}.  The rank is chosen to be the value of $R$ that gives the lowest log-likelihood of the test set.
The above approach works well in simulation studies, as shown in Appendix~\ref{appRank}.

\section{Special Cases}
\label{othermeth}
The SupCP model specified in Section~\ref{sec:mod} reduces to traditional factor analysis without supervision ($\B=\mathbf{0}$) and when $K=1$, i.e., when $\XX$ is a matrix. More generally, when $K=1$ SupCP reduces to the SupSVD model \citep{li2016supervised}.  Thus, an alternative strategy of decomposing $\XX$ is to apply SupSVD to the matricized data  $\X^{(1)}$.  However, this strategy is inefficient and therefore less accurate when the data have multi-linear structure, as shown in Sections~\ref{sims} and~\ref{app};  moreover, the patterns within each mode ($\V_k$) that are provided by a multiway factorization framework are often useful to interpret.  Lastly, the identifiability results in Section~\ref{ident} are not applicable to matrices; the SupSVD model requires additional restrictions of orthogonality and diagonal covariance for identifiability.

Without the presence of covariates $\Y$, SupCP reduces to a probabilistic CP factorization in which factors for the first mode (samples) are considered random.  The EM algorithm in Section~\ref{est} fits such a model when $\Y=\mathbf{0}$.  To our knowledge this particular unsupervised model and estimation approach, which provides a generative model for the sampled mode, is novel.  Related maximum-likelihood based CP factorizations have been proposed by \cite{mayekawa1987maximum} and \cite{vega2003maximum}, and there is a large body of work on Bayesian models for the CP and other multiway factorizations (see, e.g., \cite{hoff2011separable} and \cite{zhou2015bayesian}).

If the entries of the residual factor covariance $\Sigma_f$ are large relative to the noise variance $\sigma_e^2$, then $\U, \V_1,\hdots,\V_K$ reduce to a standard unsupervised least-squares CP factorization.  Specifically, if $\Sigma_f$ is diagonal and
\[\mbox{min}_{r} \Sigma_f[r,r] / \sigma^2_e \rightarrow \infty,\] it is easy to see that the joint likelihood (\ref{eqJointLike}) is maximized by the solution that minimizes the squared residuals for the entries of $\XX$;  the supervision component $\Y \B$ is given by a regression of $\Y$ and $\U$, but will not influence the factorization of $\XX$.

If the entries of the residual factor covariance $\Sigma_f$ are small,
\[\|\Sigma_f\|_F^2/\sigma_e^2 \rightarrow 0,\]
then the residuals $\F$ also tend to zero and the underlying structure of $\XX$ is given entirely by $\Y$:
\begin{align}
\XX = \tp{\mathbf{Y B}, \V_1, \hdots, \V_k} +  \EE.
\end{align}
This can be considered a multi-linear regression model for a multiway response ($\XX$) from vector valued predictors $\Y$.  This has received considerably less attention than the reverse scenario, prediction of a vector outcome from multiway predictors.  However, recently \cite{li2017parsimonious} proposed a tensor response regression, wherein the tensor response is assumed to have a Tucker factorization with weights determined by vector-valued predictors.  \cite{hoff2015multilinear}, describes a general framework for tensor-on-tensor regression, in which the predictor tensor and the outcome tensor have the same number of modes.

\subsection{Model complexity}
\label{complexity}

An important advantage for using a multiway factorization for multiway data, rather than vectorizing the data and using a matrix factorization, is the decrease in model complexity.  Any SupCP model is also included in the support of the SupSVD model for the matricized data, as seen by the matrix form of the model (\ref{eqSupCPmat}).  However, model complexity scales with the sum of dimensions in each mode under SupCP and with the product of dimensions in each mode under SupSVD.  The total number of free parameters in the SupCP model with diagonal covariance, accounting for the identifiability conditions of Theorem~\ref{thm1}, is
\[R \left(1+q+d_1+\cdots+d_K-K\right)+1.\] The number of free parameters for the SupSVD model, accounting for orthogonality conditions required for identifiability,  is
\[R \left(1+q+d_1d_2 \cdots d_K -(R+1)/2\right)+1.\]
Thus, model complexity for the same rank can decrease by several orders of magnitude
under SupCP,  especially for data with several high-dimensional modes.  Moreover, incorporating supervision into the likelihood model increases the model complexity by a factor of only $R q$, which is negligible  if the size of $\XX$ is much larger than the size of $\Y$.  

\section{Simulation}
\label{sims}

Here we describe a comprehensive simulation study which compares three different methods: the proposed SupCP method, the least-squares CP factorization, and the SupSVD method (on matricized data along the first mode).
We simulate data from different settings, and evaluate the performance of different methods in terms of parameter estimation and low-rank signal recovery accuracy. To avoid complication, we set the ranks for different methods to be the true ranks. The simulation study for rank estimation is conducted separately in Appendix \ref{appRank}.

\subsection{Simulation Settings}
\label{SimSettings}

We  first consider a 3-way array $\XX$ with $n=100$ samples, and $d_1=10$ and $d_2=10$ variables in the other two modes respectively.
In particular, the generative model for $\XX$ is $\XX = \tp{\U,\V_1,\V_2} + \EE$, where $\tp{\U,\V_1,\V_2}$ is the underlying low-rank signal with true rank $R=5$, and $\EE$ is normally distributed noise with iid entries with mean 0 and variance $\sigma_e^2=4$.
The auxiliary data matrix $\Y$ contains $q=10$ variables, potentially related to the latent variables in $\U$ through a linear relation $\U=\Y\B+\F$.
The coefficient $\B$ is a $10\times 5$ matrix, and the random matrix $\F$ has iid rows with mean $\mathbf{0}$ and covariance $\Sigma_f$.

We particularly consider the following settings:
\begin{itemize}
  \item {\bf Setting 1 (Non-supervision Setting)}: We set $\B$ to be a zero matrix, indicating the auxiliary data $\Y$ have no effect on the underlying structure of $\XX$ (i.e., $\XX$ is generated from a probabilistic CP model). 
      The covariance matrix $\Sigma_f$ is a diagonal matrix with diagonal values $\{25,16,9,4,1\}$.
  \item {\bf Setting 2 (Mixed Setting)}:  The coefficient matrix $\B$ is filled with Gaussian random numbers. The covariance matrix $\Sigma_f$ is the same as in Setting 1. Correspondingly, the latent score $\U$ is jointly affected by $\Y$ and $\F$.
  \item {\bf Setting 3 (Full-supervision Setting)}: The coefficient matrix $\B$ is the same as in Setting 2, and the covariance matrix $\Sigma_f$ is a zero matrix. Namely, the latent score $\U$ is solely determined by $\Y$.
\end{itemize}
In all of the above settings, the loading matrices $\V_1$ and $\V_2$ are filled with random numbers and normalized to have orthonormal columns. We also consider several additional simulation settings in Appendix \ref{add:sim}, where 1) loadings are highly collinear; 2) dimensions $d_1$ and $d_2$ are high; 3) there are more than 3 modes; 4) the model is misspecified and data do not have multiway structure.

We conduct 100 simulation runs for each setting, and compare different methods using various criteria. In particular, to evaluate the overall performance of dimension reduction, we assess the Frobenius norm of the difference between the estimated and true low-rank tensors $\|\tp{\widehat{\U},\widehat{\V}_1,\widehat{\V}_2} -\tp{\U,\V_1,\V_2} \|_F$ (denoted by $SE$).
We also consider the estimation accuracy of loadings via the maximal principal angles $\angle(\V_1,\widehat{\V}_1)$ and $\angle(\V_2,\widehat{\V}_2)$  \citep{bjorck1973numerical}. This only applies to SupCP and CP because SupSVD decomposes $\X^{(1)}$ and there is no direct estimation of $\V_1$ and $\V_2$.
In addition, we evaluate the estimation accuracy of other important parameters in SupCP and SupSVD via $\|\B-\widehat{\B}\|_F$, $\left(\sigma_e^2-\widehat{\sigma_e^2}\right)/\sigma_e^2$ (relative error, denoted by $RE_e$), and the mean relative errors of the diagonal values of $\Sigma_f$ (denoted by $RE_{f}$).
We also compare the fitting times of different methods on a standard desktop computer (8Gb Ram, 3.30GHz).

\subsection{Results}\label{sim:result}

The simulation results for Settings 1--3 are presented in Table \ref{tab:sim}. Additional results can be found in Appendix \ref{add:sim}.
In all settings, SupCP always provides the smallest $SE$.
Namely, it has the highest low-rank signal estimation accuracy.
This is true even in Setting 1 where the data are generated from a CP model.
We remark that SupCP outperforms CP in the non-supervision setting due to the shrinkage estimate of $\U$ in \eqref{Uhat} from the EM algorithm.
The shrinkage estimate strikes a balance between bias and variance, and thus leads to a smaller $SE$.
A similar result was reported in \cite{li2016supervised}.
From the dimension reduction point of view, SupCP outperforms the competing methods and identifies the most accurate underlying structure.

In addition, in Setting 1, SupCP has similar loading estimation losses to CP. Namely, the proposed method  automatically approximates the unsupervised method when the auxiliary data are irrelevant.
SupCP also outperforms SupSVD in terms of the estimation of other parameters.
In Setting 2, SupCP almost uniformly outperforms the competing methods.
In Setting 3, where the latent score is solely determined by the auxiliary data, both supervision methods (SupCP and SupSVD) outperform CP in the signal recovery accuracy. Moreover, SupCP significantly improves the loading estimation over CP. 
In terms of the fitting time, CP is the fastest while SupCP and SupSVD are also very efficient in these settings.
However, when dimensions are higher or the number of modes is larger, the computation of the matrix-based method SupSVD quickly becomes infeasible.
Both SupCP and CP scale well to larger data sets (see Appendix \ref{add:sim}).
Overall, SupCP is computationally efficient and has a good performance in a range of settings.

\begin{table}[!h]
\caption{Simulation results under Settings 1, 2 and 3 (each with 100 simulation runs). The median and median absolute deviation (MAD) of each criterion for each method are shown in the table. The best results are highlighted in bold.}
\label{tab:sim}
\begin{center}
{\footnotesize
\begin{tabular}{|l|c||c|c|c|}
\hline
Setting & Criterion &  SupCP & CP & SupSVD \\
\hline
 \multirow{7}{*}{1: Non-supervision}
       & $SE$    &    {\bf 45.97 (1.38)} & 58.75 (2.37) & 66.64 (1.08)\\
       & $\angle(\V_1,\widehat{\V_1})$ & {74.93 (10.93)} & {\bf 74.23 (8.89)} & \\
       & $\angle(\V_2,\widehat{\V_2})$ & {71.30 (10.40)} & {\bf 70.11 (9.42)} &\\
       & $\|\B-\widehat{\B}\|_F$ & {\bf 34.27 (1.95)} & & {39.50 (2.48)}\\
       & $100*RE_e$ & {\bf 1.75 (0.84)} & & 7.85 (0.96)\\
       & $100*RE_f$ & {\bf 37.66 (14.67)}&  & {143.83 (26.77)}\\
       & Time & 0.36 (0.09) & {\bf 0.06 (0.02)}& 0.24 (0.09)\\

\hline

 \multirow{7}{*}{2: Mixed}
       & $SE$    & {\bf 42.45 (0.97)}& 51.83 (1.82)& 61.20 (0.92)\\
       & $\angle(\V_1,\widehat{\V_1})$ & {\bf 10.58 (1.64)}& 12.34 (3.07) &\\
       & $\angle(\V_2,\widehat{\V_2})$ & {\bf 10.94 (1.63)} & 13.21 (3.17)&\\
       & $\|\B-\widehat{\B}\|_F$ &{\bf 31.51 (1.81)}&& 48.61 (6.30)\\
       & $100*RE_e$ &{\bf 1.29 (0.80)}&&6.10 (0.83)\\
       & $100*RE_f$ &{\bf 24.00 (6.06)}&&{ 30.28 (9.32)}\\
       & Time &0.32 (0.06)& {\bf 0.04 (0.02)}& 0.23 (0.07)\\

\hline

 \multirow{7}{*}{3: Full-supervision}
       & $SE$    & {\bf 25.06 (1.04)}& 53.95 (1.98)& 53.23 (1.06)\\
       & $\angle(\V_1,\widehat{\V_1})$ &{\bf 12.88 (1.78)}& 18.16 (4.85)&\\
       & $\angle(\V_2,\widehat{\V_2})$ &{\bf 12.99 (1.63)}& 17.67 (4.35)&\\
       & $\|\B-\widehat{\B}\|_F$ & 120.44 (11.63)&& {\bf 109.02 (7.24)}\\
       & $100*RE_e$ &{\bf 1.77 (0.89)}&& 7.40 (0.87)\\
       & $100*RE_f$ &NA&&NA\\
       & Time &0.85 (0.08)& {\bf 0.04 (0.02)}& 0.70 (0.27)\\
\hline
\end{tabular}
}
\end{center}
\end{table}

\section{Application to Amino Acid Fluorescence}
\label{appAmino}

We consider fluorescence data for five laboratory samples, measured over the emission and excitation frequency domains with a spectral fluorometer.   Intensities are available on a 2D grid for excitation wavelengths between 250nm and 450nm, and emission wavelengths   between 250nm and 310nm, in increments of 1 nm.  Thus, the resulting fluorescence array is of dimension $\XX: 5 \times 61 \times 201$.  Each sample is comprised of a mixture of three amino acids diluted in water: Tryptphan, Tyrosine, and Phenylalanin.  The known concentration of each amino acid in Mole/L is given in the matrix $\Y: 5 \times 3$.  These data have been previously published \cite{bro1998multi} and are freely available online at \url{http://www.models.life.ku.dk/nwaydata} (accessed 02/28/2018). 

\begin{figure}[!h]
\centering
   \includegraphics[scale=0.48, trim={1cm 0.5cm 1cm 0.5cm}, clip=TRUE]{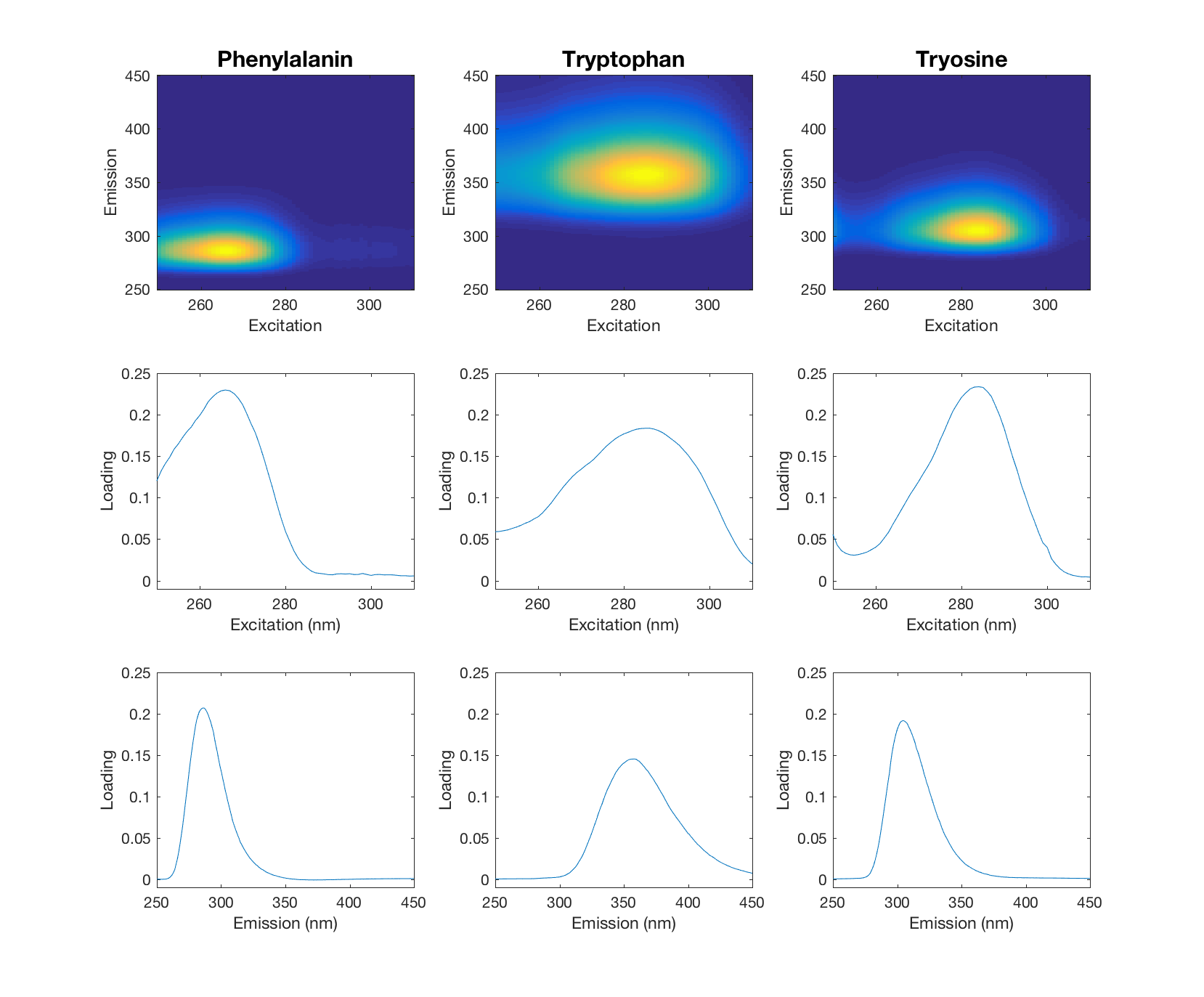}
\caption{Rank-1 components corresponding to Phenylalin, Tryptophan and Tryosine.  The bottom two rows give the loadings for emission and excitation wavelengths, and the top row gives their resulting product. }
\label{fig:amino}
\end{figure}

The structure of excitation/emission fluorescence data is suitably characterized by a CP factorization \cite{andersen2003practical}, to decompose latent patterns in the excitation and emission domains.  We apply SupCP to the fluorescence array $\XX$, supervising on $\Y$ to capture the relationship between fluorescence structure and the three amino acids.  A rank$-3$ model is selected to maximize the test likelihood (Section~\ref{rankSelect}) under a leave-one-out cross validation scheme. The resulting supervision weights $\B$ are given in Table~\ref{tabamino}, scaled by the standard deviation of each amino acid.  These results clearly show that each of the three components are driven by the concentration of a different amino acid.  The resulting components are shown in Figure~\ref{fig:amino}, and can be interpreted as the excitation and emission fluorescence spectra that are specific to each amino acid. 

  \begin{table}[!h]
\caption{Scaled coefficients for rank$-3$ fluorescence model.}
  \label{tabamino}
\begin{center}
\footnotesize
\begin{tabular}{|l|c|c|c|}
\hline
 & Component $1$ & Component $2$ & Component $3$ \\
\hline
$B_{\text{Phe}}*\mbox{sd}(Y_{\text{Phe}})$    &  \textbf{7638}  & 138 & 130 \\
$B_{\text{Trp}}*\mbox{sd}(Y_{\text{Trp}})$     & 140 & \textbf{11734} & -5\\
$B_{\text{Tyr}}*\mbox{sd}(Y_{\text{Tyr}})$        & 87 & -72 & \textbf{8514} \\
\hline
\end{tabular}
\end{center}
\end{table}

\section{Application to Facial Images}
\label{app}

The Labeled Faces in the Wild database \citep{learned2016labeled} is a publicly available set of over $13000$ images, where each image includes the face of an individual.  The images are taken from the internet and the individuals represented are celebrities.  Each image is labeled with the name of the individual, but the images are unposed and exhibit wide variation in lighting, image quality, angle, etc.\ (hence ``in the wild").

\cite{kumar2009attribute} developed an attribute classifier, which gives describable attributes for a given facial image.  These attributes include characteristics that describe the individual (e.g., gender, race, age), that describe their expression (e.g., smiling, frowning, eyes open), and that describe their accessories (e.g., glasses, make-up, jewelry).  These attribute were determined on the Faces in the Wild dataset, as well as other facial image databases.  In total $72$ attributes are measured for each image.  Our goal is to use these characteristics to supervise dimension reduction of the images and to develop a probabilistic model for computationally generating faces with given describable characteristics.

PCA and other low-rank matrix factorization approaches are commonly used to analyze facial image data.  For example, the terminology \emph{eigenfaces} \citep{sirovich1987low,turk1991eigenfaces} has been used to describe a broad class of methods in which PCA of vectorized facial images precedes a given task, such as face recognition.  PCA can only be applied to vectorized images, wherein $d_1 \times d_2$ dimensional images are converted to a $d_1 \cdot d_2$ dimensional vectors (or $d_1 \cdot d_2 \cdot 3$ dimensional vectors for color images).  Thus, PCA does not exploit multiway structure.  Although facial images are not obviously multi-linear, the CP factorization has been shown to be much more efficient as a dimension reduction tool for facial images than PCA, and marginally more efficient than other multiway factorization techniques \citep{lockComment}.

\begin{figure}[!h]
\centering
  \subfigure[Michael Jordan]{
        \includegraphics[scale=0.71]{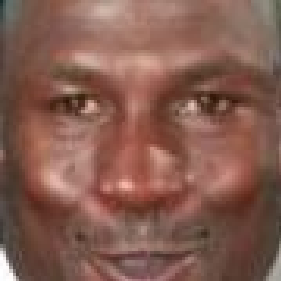}}
  \subfigure[George W. Bush]{
        \includegraphics[scale=0.71]{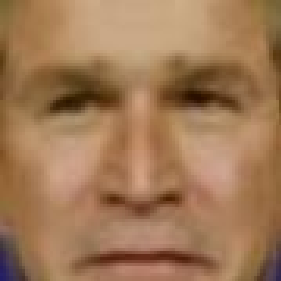}}
   \subfigure[Angelina Jolie]{
        \includegraphics[scale=0.71]{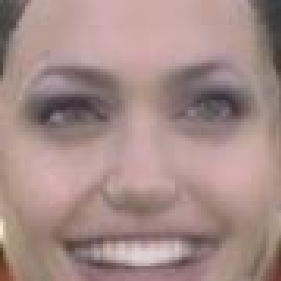}}
\caption{Examples of frontalized faces.}
\label{fig:ExampFaces}
\end{figure}

For our application the images are first \emph{frontalized}, as described in \cite{hassner2015effective}.  That is, the unconstrained images are rotated, scaled, and cropped so that all faces appear forward-facing and the image shows only the face.  After this processing step, we expect the nose, mouth and other facial features to be in approximately the same location across the images.  This step is important for linear factorization approaches such as PCA or SupCP, which assume the coordinates are consistent across the images.  Some examples of the frontalized images are shown in Figure~\ref{fig:ExampFaces}.

We work with a random sample of $4000$ frontalized images from unique individuals.  Each image is $90 \times 90$ pixels, and each pixel gives the intensity for colors red, green and blue,  resulting in a multiway array of dimensions $\XX: 4000 \times 90 \times 90 \times 3$.  We center the array by subtracting the ``mean face" from each image, i.e., we center each pixel triplet ($x \times y \times$color) to have mean $0$ over the $4000$ samples.   The attribute matrix $\Y: 4000 \times 72$  is measured on a continuous scale;  for example, for the smiling attribute, higher values correspond to a more obvious smile and lower values correspond to no smile.  We standardize $\Y$ by subtracting the mean and dividing by the standard deviation for each row, hence converting each attribute to its z-scores.

\begin{figure}[!h]
\centering
\subfigure[SupCP]{
        \includegraphics[scale=0.50]{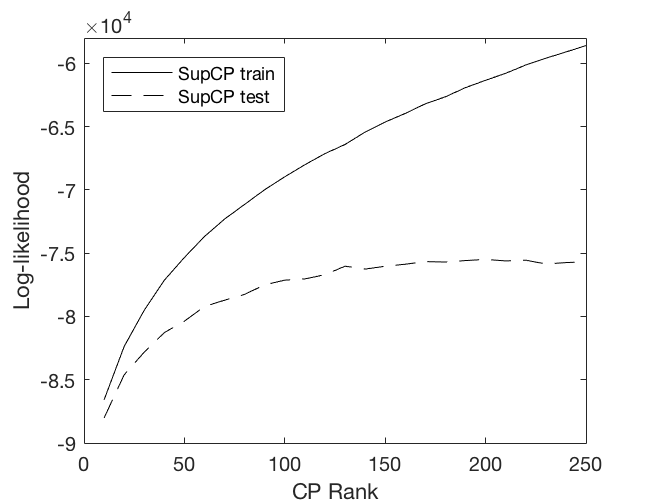}}
  \subfigure[SupSVD]{
        \includegraphics[scale=0.50]{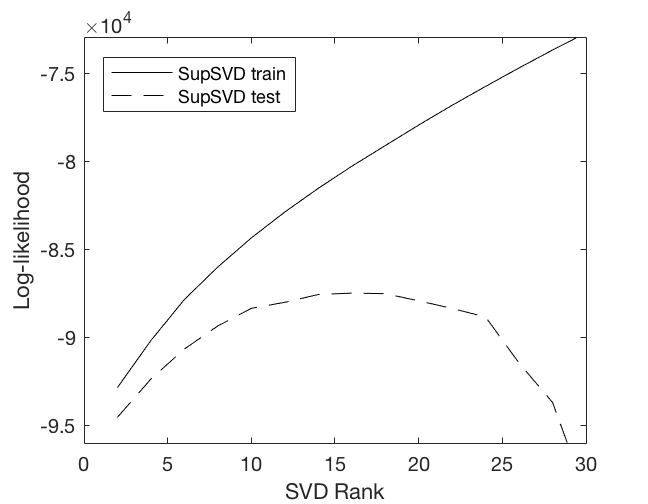}}
\caption{Log-likelihood for the estimated SupCP and SupSVD models, for different ranks, on the training images used to fit the model and the test images.}
\label{facesCV}
\end{figure}

We assess the fit of the SupCP model with a training set of $100$ randomly sampled images.  The mean log-likelihood for this training set and the mean log-likelihood for the remaining test images are shown for different ranks in Figure~\ref{facesCV}(a).  The log-likelihood increases consistently for the training set, but the test log-likelihood peaks at a rank of $200$, suggesting that over-fitting decreases the generalizability of the estimates for higher ranks.  For comparison, we also assess the SupSVD model on the vectorized images $\X^{(1)}: 4000 \times 24300$ in Figure~\ref{facesCV}(b).  The SupSVD model does not perform as well, achieving a maximum mean log-likelihood of $-8.75 \times 10^4$, versus a maximum of  $-7.55 \times 10^4$ for the SupCP model.  Note that SupSVD requires many more parameters for each component, on the order of $90 \cdot 90 \cdot 3 = 24300$ parameters per component versus $90+90+3 = 183$ for SupCP.    Thus, the SupSVD model is more prone to overfitting with a smaller number of components, which is inadequate to capture the complex covariance structure of the images and the covariate effects. A factorization using SupCP with no covariates has lower test log-likelihood for all ranks considered and achieves a maximum of $-7.60 \times 10^4$.

\begin{figure}[!h]
\centering
\subfigure[Annealing]{
        \includegraphics[scale=0.50]{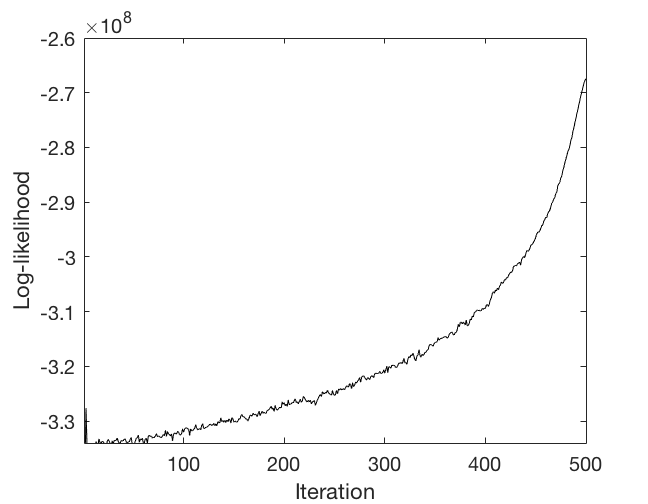}}
  \subfigure[After annealing]{
        \includegraphics[scale=0.50]{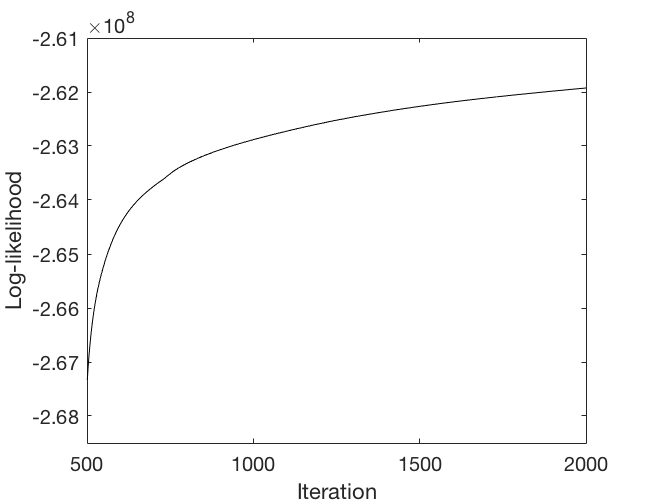}}
\caption{Log-likelihood for the model over $2000$ EM iterations.  The first $500$ iterations (a) incorporate random annealing to avoid local modes.}
\label{LikePlot}
\end{figure}

We fit the SupCP model, with rank $200$, on the full set of $4000$ images.  This job required a computing server with 64GB of RAM.  We  run the EM algorithm for $2000$ iterations, including a $500$ annealing burn-in, which took approximately $8$ hours.  After $2000$ iterations the log-likelihood appears to converge, as shown in Figure~\ref{LikePlot}.  The estimated parameters provide a data-driven generative model for faces with given covariates.  In particular, given a feature vector $\mathbf{y}$ with scores for the $72$ describable attributes, the ``mean face" for the given attributes is
\[\tp{ \mathbf{y}^T\B, \V_1,\V_2,\V_3},\]
where $\V_1: 90 \times 200$ contains loadings in the $x$-dimension, $\V_2: 90 \times 200$ contains loadings in the $y$-dimension, and $\V_3: 3 \times 200$ contains color loadings.

\begin{figure}[!h]
\centering
  \subfigure[Mean]{
        \includegraphics[scale=0.71]{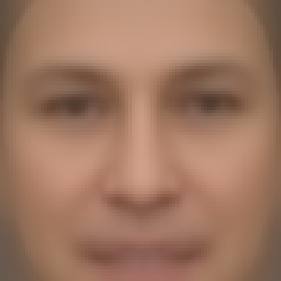}}
  \subfigure[Male]{
        \includegraphics[scale=0.71]{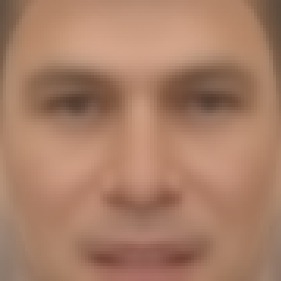}}
   \subfigure[Male; moustache]{
        \includegraphics[scale=0.71]{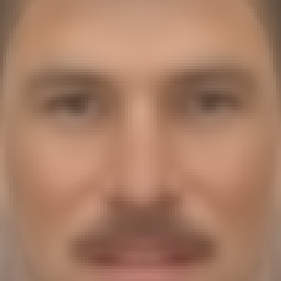}}
   \subfigure[Male; asian]{
        \includegraphics[scale=0.71]{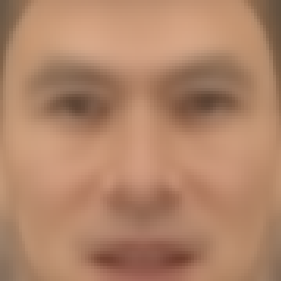}}
   \subfigure[Male; smiling]{
        \includegraphics[scale=0.71]{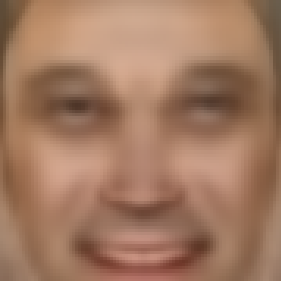}}
    \subfigure[Female; black; lipstick]{
        \includegraphics[scale=0.71]{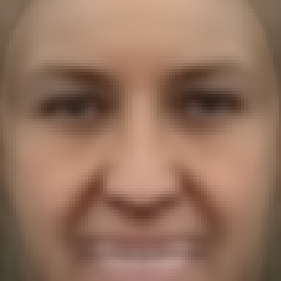}}
\caption{Constructed faces under different attribute covariates: SupCP.}
\label{fig:consFaces}
\end{figure}

\begin{figure}[!h]
\centering
  \subfigure[Mean]{
        \includegraphics[scale=0.71]{MeanFig.png}}
  \subfigure[Male]{
        \includegraphics[scale=0.71]{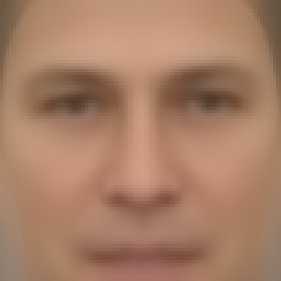}}
   \subfigure[Male; moustache]{
        \includegraphics[scale=0.71]{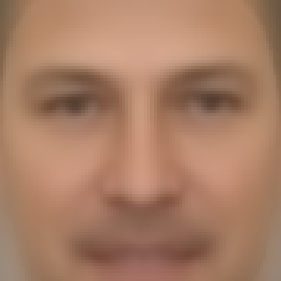}}
   \subfigure[Male; asian]{
        \includegraphics[scale=0.71]{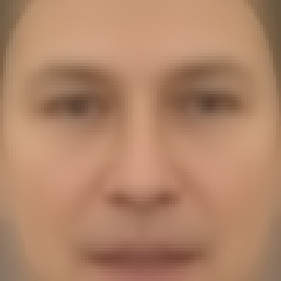}}
   \subfigure[Male; smiling]{
        \includegraphics[scale=0.71]{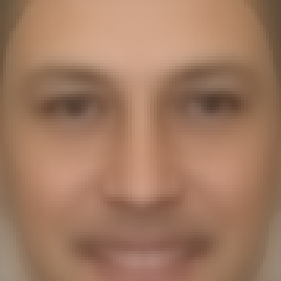}}
    \subfigure[Female; black; lipstick]{
        \includegraphics[scale=0.71]{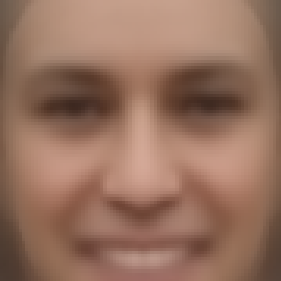}}
\caption{Constructed faces under different attribute covariates: SupSVD.}
\label{fig:consFacesSVD}
\end{figure}

The constructed mean facial images for certain attributes are given in Figure~\ref{fig:consFaces}.  To generate these images, the presence of a given features is coded as a z-score of $3$ in the covariate vector, and all other scores are set to $0$.  The resulting images are smooth and remarkably intuitive.  For comparison the analogous constructed images using SupSVD with rank $16$ are shown in Figure~\ref{fig:consFacesSVD}, and the features are less distinctive.

\newpage
\section{Discussion}
\label{s:discuss}
High-dimensional data with multiway structure are often encountered in modern research, yet present statistical methods to deal with such data are in their infancy and leave  much room for further development.  Most of the methodological development for the analysis of multiway data has traditionally come from fields other than statistics, such as computer science and chemometrics. Perhaps in part because of this, methods for multiway data are often not based on a generative statistical model.  In this article we have developed the SupCP method, which is based on a probabilistic CP factorization model in which additional covariates inform the factorization.  Our simulation studies illustrate the advantages of SupCP for dimension reduction and interpretation when the underlying structure of multiway data is multi-linear and at least partially driven by additional covariates.  Our application to amino acid data illustrates how supervision can facilitate interpretation of the underlying low-rank components. Our application to facial image data illustrates how SupCP can be used to more generally estimate a generative model for multiway data with given covariates.

Many datasets can be represented as multiway arrays, but we caution that SupCP is not well motivated if the data do not have any multi-linear structure.  In particular, SupCP and other approaches that involve multiway factorization are not universally applicable to visual image analysis problems.  For our application to facial analysis it is critical that the images are well-aligned to common coordinates, and share common structure on that coordinate system. The data also show some (high-rank) multiway structure, as SupCP is more accurate than the analogous approach that relies on vectorization of the multiway images.  State-of-the-art generative deep learning models, such as a properly trained conditional variational auto encoder, have also been applied to the task of facial image generation with promising results \cite{yan2016attribute2image}.

Our model assumes that array data can be decomposed into low-rank structure and independent residual error.  Extensions to the model may allow for more complex residual correlation structures, such as a separable covariance for a more general array normal framework \cite{hoff2011separable}.

The SupCP framework introduced in this article may be extended in several ways.  Here we have focused on the CP factorization because it allows for a straightforward extension of the supervised SVD model and has been shown to be an efficient factorization for facial image data.  Supervised and probabilistic approaches to the Tucker factorization, or other multiway dimension reduction techniques, are interesting directions for future work.  Moreover, there is a growing body of work on multiway factorization methods for sparse or functional data \citep{allen2012,allen2013multi}.  The SupCP framework may be extended to accommodate sparse and functional data, or non-linear covariate effects, which are analogous to recent extensions of the SupSVD framework \citep{li2015supervised,fan2016projected}.   Finally, here we have focused on allowing a random latent variable and covariate supervision for one mode (the sample mode).  More general approaches which allow for supervision or random components in more than one mode is another promising direction of future work.

\appendix

\section{Proof of Theorem~\ref{thm1}}
\label{app1}

Here we show that the parameters $\{\V_1,\hdots,\V_K, \Sigma_f, \sigma^2_e, \B\}$ are identifiable given the likelihood~(\ref{eqLike}), for the conditions of Theorem~\ref{thm1}.  By the identifiability of the mean and covariance of a multivariate normal likelihood, it suffices to show that any given mean
\[\mu_\X = \Y \B\Vmat^T \]
and covariance
\[\Sigma_\X = \Vmat \Sigma_f \Vmat^T+\sigma^2_e \mathbf{I}_d\]
uniquely identify $\V_1,\hdots, \V_K$, $\Sigma_f$, $\sigma^2_e$, and $\B$.

Note the entries of $\mu_\X$ can be re-arranged as the (K+1)-way array \[\tilde{\mu}_\XX = \Y \B \circ \V_1 \circ \hdots \circ \V_K\]
of dimensions $n \times d_1 \times \cdots \times d_K$.  By the result of \cite{sidiropoulos2000uniqueness}, the components $\{\Y,\B, \V_1, \hdots \V_K\}$ are uniquely defined up to scaling and ordering if the $k$-rank condition \eqref{krCond} is satisfied.  Hence, by the scaling and ordering restrictions (a) and (b), the components are fully identified.  Also, the condition that $\Y$ is of full column rank assures that $\B$ is uniquely identified by
\[\B = (\Y^T \Y)^{-1} \Y^T \U.\]
Further, $\sigma_e^2$ is uniquely identified as the value that gives \[\mbox{rank} \left(\Sigma_\X - \sigma^2_e \I_d \right) = R.\]
Letting
\[\Sigma_R = \Sigma_\X - \sigma^2_e \I_d = \Vmat \Sigma_f \Vmat^T,\]
we see that $\Sigma_f$ is uniquely identified by
\[\Sigma_f =  (\Vmat^T \Vmat)^{-1} \Vmat^T \Sigma_R \Vmat (\Vmat^T \Vmat)^{-1}.  \] $\hfill$ $\square$

\section{M-step derivations}
\label{app3}
Here we give details for the conditional maximum likelihood estimates given in Section~\ref{Mstep}.

To update $\V_1$, we fix $\V_2,\hdots,\V_K$ and set the partial derivative of the log-likelihood with respect to $\V_1$ equal to $0$.  It helps to define $\W_{\text{mat}}^{(1)}:\left( n \prod_{k=2}^K d_k\right) \times R$ as the design matrix for regressing $\X^{(2)}$ on $\V_1$. Given $\mathbf{u}_r$ is the $r$th column of $\U$, the $r$th column of $\W_{\text{mat}}^{(1)}$ is the vectorization of $\mathbf{u}_{r} \circ \v_{2r} \circ \cdots \circ \v_{Kr}.$  Then, \[\X^{(2)} = \V_1 \W_{\text{mat}}^{(1)} + \E^{(2)},\]
and the derivative of the joint log-likelihood with $L(\XX, \U)$ with respect to $\V_1$ is
\begin{align} \label{Vpartial}
\frac{\partial L(\XX, \U)}{\partial \V_1} = 2 \sigma^2_e \left(\X^{(2)^T} \W_{\text{mat}}^{(1)} - \V_1 \W_{\text{mat}}^{(1)^T} \W_{\text{mat}}^{(1)} \right).
 \end{align}
Let $\widehat{\W}_{\text{mat}}^{(1)}$ be the conditional expectation of $\W_{\text{mat}}^{(1)}$.  Then, the $r$th column of $\widehat{\W}_{\text{mat}}^{(1)}$ is the vectorization of $\widehat{\mathbf{u}}_{r} \circ \v_{2r} \circ \cdots \circ \v_{Kr}$, where $\widehat{\mathbf{u}}_{r}$ is the $r$th column of $\widehat{\U}$.  Note that
\[\W_{\text{mat}}^{(1)^T} \W_{\text{mat}}^{(1)} = (\Vmat^{(1)^T} \Vmat^{(1)}) \cdot \U^T \U\]
where $\Vmat^{(1)}: \prod_{k=2}^K d_k \times R$  is the matrix with column $r$ being the vectorization of
$\v_{2r} \circ \cdots \circ \v_{2r}$.  Thus the conditional expectation of $\W_{\text{mat}}^{(1)^T} \W_{\text{mat}}^{(1)}$ is
\[(\Vmat^{(1)^T} \Vmat^{(1)}) \cdot \left(\widehat{\U}^T \widehat{\U} + n\widehat{\Sigma}_{U} \right).\]
Taking the conditional expectation of (\ref{Vpartial}) and setting equal to $0$ yields
\[\V_1 = \X^{(1)^T} \W_{\text{mat}}^{(1)} \left(\Vmat^{(1)^T} \Vmat^{(1)} \cdot (\widehat{\U}^T \widehat{\U} + n\widehat{\Sigma}_{U}) \right)^{-1}.\]
The remaining loading matrices $\V_2, \hdots, \V_K$ are updated similarly.

The partial derivatives of $\B$, $\Sigma_f$, and $\sigma^2_e$ are
\begin{align*}
\frac{\partial L(\XX, \U)}{\partial \B} &= 2(\Y^T \U - \Y^T\Y \B) \Sigma_f^{-1} \\
\frac{\partial L(\XX, \U)}{\partial \Sigma_f} &=	-n \Sigma_f^{-1}+\Sigma_f^{-1} (\U-\Y\B)^T(\U-\Y\B)\Sigma_f^{-1} \\
\frac{\partial L(\XX, \U)}{\partial \sigma^2_e} &= -n d \sigma_e^{-2} + \sigma_e^{-4} \tr \left((\X^{(1)}-\U \Vmat^T)(\X^{(1)}-\U \Vmat^T)^T \right).
\end{align*}
Taking the conditional expectation of the above partial derivatives and setting them equal to $0$ yields the parameter updates in (\ref{paramUpdates}).

\section{Rank Selection Simulation}
\label{appRank}

Here we give the results of a simulation to asses the performance of the likelihood cross-validation approach to rank selection described in Section~\ref{rankSelect}. The data simulation scheme is similar to that in Section~\ref{SimSettings}, but the rank $R$ is varied for each simulation.  Specifically, data are generated for the 3-way array $\XX: n \times d_1 \times d_2$ with $n=100$ samples, $d_1=25$ and $d_2 = 25$, and $\Y$ is an auxiliary matrix with $10$ variables, $\Y: 100 \times 10$.  Given rank $R$ structure, the latent signals are generated under model $(\ref{eqSupCP}$) as follows:
\begin{enumerate}
\item The entries of $\Y$ and $\B: 10 \times R$ are generated independently from a standard normal distribution.
\item The diagonal entries of the covariance $\Sigma_f: R \times R$ are generated from a Uniform$(5,25)$ distribution, and the factor residuals $\F: n \times R$ have iid rows with mean $0$ and covariance $\Sigma_f$. 	
\item The entries of $\V_1: 25 \times R$ and $\V_2: 25 \times R$ are generated to have orthonormal columns via the SVD of a random matrix.
\end{enumerate}

The error array $\EE: 100 \times 25 \times 25$ is generated by independent $N(0,\sigma^2)$ entries, where we consider differing noise levels given by $\sigma^2 = 1, 5, $ or $10$.  For each noise level, we independently generate $10$ datasets as above for each of $R=0,1,\hdots,10$. For each of the $3 \cdot 10 \cdot 11 = 330$ simulated datasets, we estimate the rank via the likelihood cross-validation scheme described in Section~\ref{rankSelect}.  We randomly specify a training set and a test set, each of size $50$, and consider the ranks $r=0,1,2,\hdots,9,$ or $10$,  choosing the rank that gives the lowest log-likelihood in the test set.   Note that $R=0$ corresponds to a null dataset with independent entries, and the results for $r=0$ are given by a the independent normal log-likelihood of the test set with variance $\hat{\sigma}^2$  given by the maximum likelihood estimate of the training entries.

\begin{figure}[!h]
\centering
\includegraphics[width=1\textwidth, trim={5cm 00cm 4cm 0.2cm}, clip=TRUE]{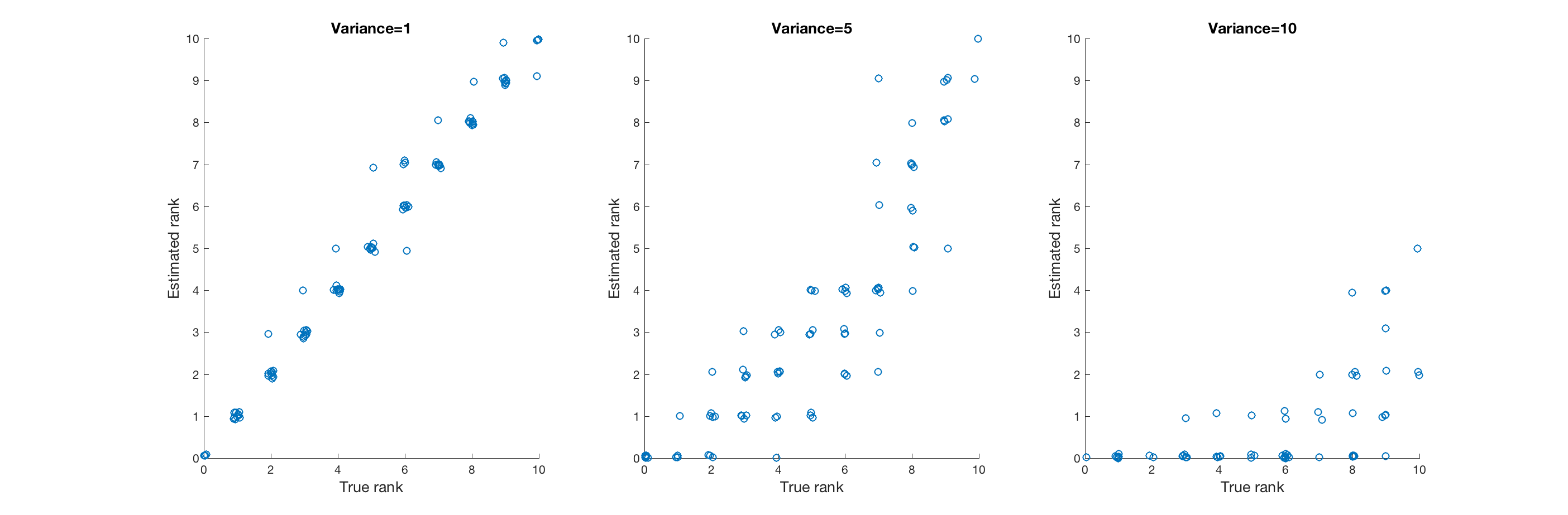}
\caption{True rank vs. estimated rank for $100$ randomly generated simulations, under different levels of noise variance.  A small amount of jitter is added to show multiple points with the same coordinates.}
\label{rankFig}
\end{figure}

Figure~\ref{rankFig} shows a scatterplot of the estimated rank and the true rank for each noise level.  For $\sigma^2=1$ the estimated rank is a good predictor for the true rank;   the correct ranks were chosen in 83 of the  simulations, overestimated slightly in 13, and underestimated slightly for 4.  This demonstrates that a likelihood cross-validation approach is reasonable to determine the rank of the model.  For $\sigma^2=5$, the estimated rank still shows a clear association with the true rank, but it is less accurate, with a tendency toward under-estimation.  When $\sigma^2=10$ the ranks are severely underestimated, as the low-rank signal is difficult to distinguish from the noise.  For all null simulations ($R=0$) the approach correctly does not identify any signal.

\section{Initialization Simulation}
\label{app2}

Here we describe the results of a simulation to assess the sensitivity of the iterative algorithm in Section~\ref{EMalg} to initial values.  The algorithm can converge to a local maximum, and we explore different strategies to alleviate this issue.

Our simulation scheme is as follows.  We consider a $4$-way array $\XX$ of dimensions  $10 \times 20 \times 40 \times 50$, and a vector of covariates $\y: 10 \times 1$. For rank $R=2$,
\begin{enumerate}
\item The entries of $\U: 2 \times 10$ are generated independently from a normal distribution, variance of each row chosen from Uniform$(2,22)$
\item The entries of $\y$ are correlated with first column of $\U$; specifically,  $\y=\mathbf{u}_1-\mathbf{f}$ where the entries of $\mathbf{f}$ are generated from a standard normal distribution
\item The entries of $\V_1: 20 \times 2$, $\V_2: 40 \times 2$, $\V_3: 50 \times 2$  are generated independently from a standard normal distribution and standardized so that the columns have unit norm.
\item The error array $\EE: 10 \times 20 \times 40 \times 50$ is given by independent $N(0,1)$ entries.
\end{enumerate}

We generate $100$ datasets under the above scheme, and for each dataset we run the algorithm under different initialization methods.  We apply both random initialization and initialization via a CP decomposition, as described in Section~\ref{init}.  To avoid local maximum, we also consider a form of annealing in which random variation is incorporated during the first few iterations.  Specifically, independent normally distributed noise with mean $0$ is added to the expected value for $\U$ at each iteration (Section~\ref{Estep}), for the first $L$ iterations, and the standard deviation of the noise is proportional to the inverse of the iteration number.

In total we consider six initialization methods: (1) random initialization with no annealing, (2) random initialization with $100$ annealing iterations, (3) random initialization with $500$ annealing iterations, (4) CP initialization with no annealing, (5) CP initialization with $100$ annealing iterations, and (6) CP initialization with $500$ annealing iterations.  For each method, and each simulated data set, we run the algorithm until convergence twice under different random seeds.

In Table \ref{tab:init} we show the mean absolute difference in log-likelihoods between different runs, and the mean likelihood, under the different methods.  Under this scheme, a CP initialization generally gives little benefit over a random initialization, and annealing improves the log-likelihood after convergence and the agreement under different runs. However, even $500$ annealing iteration did not result in perfect agreement between different runs of the algorithm.  To further reduce convergence to local optimum, one can run the algorithm under multiple initialization and choose that which gives the highest log-likelihood after convergence.

\begin{table}[!h]
\caption{Results under 6 different initialization approaches over $100$ simulated datasets.  The mean log-likelihood, mean absolute difference between runs for the same data, and mean computing time for a single run are shown, with standard deviations in parentheses.   }
\label{tab:init}
\begin{center}
{\footnotesize
\begin{tabular}{|l|c|c|c|}
\hline
Method & Log-likelihood & Difference & Time (sec) \\
\hline

Random start, no annealing     & -200160 (589)   & 297.6 (576) & 3.6 (3.1)\\
Random start, $100$ annealing iters      & -200090 (529) & 180.8 (410) & 4.7  (3.2)\\
Random start, $500$ annealing iters        & -199980 (420) & 58.2 (132) & 9.0 (2.6) \\
CP start, no annealing       & -200320 (752) & 433.5 (743) & 3.1 (2.6)\\
CP start, $100$ annealing iters       & -200080 (501) & 183.1 (404) & 5.1 (3.3)\\
CP start, $500$ annealing iters       & -199980 (422) &53.6 (149)  & 9.6 (2.6)\\
\hline
\end{tabular}
}
\end{center}
\end{table}

\section{Additional Simulation Results}\label{add:sim}
\subsection{Collinearity Simulation}
Here we investigate the effect of loading collinearity on model estimation. We conduct additional simulation studies under Settings 1--3, while setting the loadings in $\V_1$ and $\V_2$ to be highly collinear, respectively.
In particular, we fill $\V_1$ and $\V_2$ with random numbers, and only normalize the columns to have unit norm.
The other parameters are kept unchanged from Settings 1--3 in Section \ref{sims}.

We conduct 100 simulation runs in each setting, and the results are presented in Table \ref{tab2}.
In general, the results are very similar to those in Section \ref{sim:result} (where true loadings are orthonormal). SupCP significantly outperforms CP and SupSVD in the low-rank signal estimation. In terms of the estimation of each individual parameter, SupCP is better than or at least comparable to the corresponding competing methods.
We note that principal angles $\angle(\V_1,\widehat{\V}_1)$ and $\angle(\V_2,\widehat{\V}_2)$ for SupCP and CP are generally larger and have higher variabilities under the collinear settings compared to the orthonormal settings.
This may be due to the local optimum issue of the low-rank approximation to a tensor array, which calls for further investigation.

\begin{table}[!h]
\caption{Simulation results under Settings 1, 2 and 3 with collinear loadings (each with 100 simulation runs). The median and median absolute deviation (MAD) of each criterion for each method are shown in the table. The best results are highlighted in bold.}
\label{tab2}
\begin{center}
{\footnotesize
\begin{tabular}{|l|c||c|c|c|}
\hline
Setting & Criterion &  SupCP & CP & SupSVD \\
\hline
 \multirow{7}{*}{1: Non-supervision}
       & $SE$    &    {\bf 45.28 (1.11)} & 58.60 (1.86) & 65.94 (1.08)\\
 \multirow{7}{*}{(collinear loadings)}
       & $\angle(\V_1,\widehat{\V_1})$ & {\bf 73.01 (10.09)} & {73.97 (10.07)} & \\
       & $\angle(\V_2,\widehat{\V_2})$ & {\bf 73.82 (9.53)} & {74.40 (9.56)} &\\
       & $\|\B-\widehat{\B}\|_F$ & {\bf 35.56 (1.89)} & & {41.28 (2.40)}\\
       & $100*RE_e$ & {\bf 1.89 (0.87)} & & 7.99 (0.94)\\
       & $100*RE_f$ & {\bf 38.19 (12.63)}&  & {132.92 (31.69)}\\
       & Time & 0.48 (0.14) & {\bf 0.06 (0.02)}& 0.29 (0.12)\\

\hline
 \multirow{7}{*}{2: Mixed}
       & $SE$    & {\bf 42.75 (1.11)}& 54.29 (0.91)& 61.70 (1.05)\\
 \multirow{7}{*}{(collinear loadings)}
       & $\angle(\V_1,\widehat{\V_1})$ & {\bf 55.40 (15.21)}& 68.39 (12.33) &\\
       & $\angle(\V_2,\widehat{\V_2})$ & {\bf 69.45 (10.89)} & 75.32 (10.12)&\\
       & $\|\B-\widehat{\B}\|_F$ &{\bf 52.54 (13.81)}&& 87.89 (7.39)\\
       & $100*RE_e$ &{\bf 1.40 (0.78)}&& 6.90 (0.90)\\
       & $100*RE_f$ &{\bf 29.35 (10.17)}&& {66.01 (17.27)}\\
       & Time &0.58 (0.21)& {\bf 0.05 (0.02)}& 0.24 (0.12)\\

\hline

 \multirow{7}{*}{3: Full-supervision}
       & $SE$    & {\bf 27.75 (1.29)}& 56.05 (1.43)& 53.08 (1.08)\\
 \multirow{7}{*}{(collinear loadings)}
       & $\angle(\V_1,\widehat{\V_1})$ &{\bf 61.49 (12.91)}& 73.88 (8.75)&\\
       & $\angle(\V_2,\widehat{\V_2})$ &{71.73 (9.47)}& {\bf 71.30 (9.12)}&\\
       & $\|\B-\widehat{\B}\|_F$ & 125.34 (11.17)&& {\bf 112.09 (7.22)}\\
       & $100*RE_e$ &{\bf 1.88 (1.01)}&& 7.74 (1.00)\\
       & $100*RE_f$ &NA&&NA\\
       & Time &1.19 (0.23)& {\bf 0.06 (0.02)}& 0.71 (0.27)\\
\hline
\end{tabular}
}
\end{center}
\end{table}

\subsection{High-Dimension Simulation}
We also investigate the performance of different methods in higher dimensions.
In particular, we fix the parameters ($\B$ and $\Sigma_f$) as in Setting 2, and increase the dimensions $(d_1,d_2)$ of the loading matrices from $(10,10)$ to $(50,50)$, $(100,100)$ and $(500,500)$.
We fill $\V_1$ and $\V_2$ with random numbers and normalize them to have orthonormal columns.
In order to keep the signal-to-noise ratio (SNR) unchanged from Setting 2, we adjust $\sigma_e^2$ accordingly.

The results of 100 simulation runs under different settings are presented in Table \ref{highdim}.
The fitting time for every method increases with the increasing dimension, but SupCP and CP can be fitted within a reasonable time in all settings.
However, SupSVD is computationally infeasible when $(d_1,d_2)=(500,500)$, where there are 250,000 variables in the matricized data.
For SupCP, the estimation accuracy of the low-rank signal and loading matrices improves with the increasing dimension and fixed SNR.
SupCP significantly outperforms CP regardless of the dimensions.

\begin{table}[!h]
\caption{Simulation results under Settings 2 with $(d_1,d_2)=(50,50),(100,100),(500,500)$ (each with 100 simulation runs). The median and median absolute deviation (MAD) of each criterion for each method are shown in the table. The best results are highlighted in bold.}
\label{highdim}
\begin{center}
{\footnotesize
\begin{tabular}{|l|c||c|c|c|}
\hline
Setting & Criterion &  SupCP & CP & SupSVD \\
\hline
 \multirow{7}{*}{Mixed Setting}
       & $SE$    &    {\bf 12.57 (0.23)} & 12.95 (0.47) & 45.90 (0.21)\\
 \multirow{7}{*}{$(d_1,d_2)=(50,50)$}
       & $\angle(\V_1,\widehat{\V_1})$ & {\bf 4.97 (0.31)} & {5.29 (0.58)} & \\
       & $\angle(\V_2,\widehat{\V_2})$ & {\bf 5.14 (0.40)} & {5.49 (0.83)} &\\
       & $\|\B-\widehat{\B}\|_F$ & {\bf 26.63 (0.26)} & & {34.46 (0.76)}\\
       & $100*RE_e$ & {\bf 0.26 (0.15)} & & 5.06 (0.18)\\
       & $100*RE_f$ & {\bf 12.24 (1.27)}&  & {28.39 (2.37)}\\
       & Time & 1.06 (0.06) & {\bf 0.07 (0.02)}& 1.07 (0.02)\\

\hline

 \multirow{7}{*}{Mixed Setting}
       & $SE$    & {\bf 7.72 (0.10)}& 7.84 (0.21)& 45.06 (0.10)\\
 \multirow{7}{*}{$(d_1,d_2)=(100,100)$}
       & $\angle(\V_1,\widehat{\V_1})$ & {\bf 3.64 (0.19)}& 3.90 (0.42) &\\
       & $\angle(\V_2,\widehat{\V_2})$ & {\bf 3.60 (0.20)} & 3.76 (0.41)&\\
       & $\|\B-\widehat{\B}\|_F$ &{\bf 26.48 (0.17)}&& 33.96 (0.36)\\
       & $100*RE_e$ &{\bf 0.13 (0.07)}&& 5.03 (0.08)\\
       & $100*RE_f$ &{\bf 11.96 (0.75)}&& {28.04 (1.06)}\\
       & Time &3.62 (0.18)& {\bf 0.22 (0.04)}& 11.31 (0.06)\\

\hline

 \multirow{7}{*}{Mixed Setting}
       & $SE$    & {\bf 2.99 (0.05)}& 31.40 (15.55)& NA\\
 \multirow{7}{*}{$(d_1,d_2)=(500,500)$}
       & $\angle(\V_1,\widehat{\V_1})$ &{\bf 1.67 (0.07)}& 59.30 (30.45)&\\
       & $\angle(\V_2,\widehat{\V_2})$ &{\bf 1.67 (0.07)}& {54.98 (34.81)}&\\
       & $\|\B-\widehat{\B}\|_F$ & {\bf 26.41 (0.08)}&& NA\\
       & $100*RE_e$ &{\bf 0.04 (0.03)}&& NA\\
       & $100*RE_f$ &{\bf 11.94 (0.29)}&&NA\\
       & Time &98.48 (17.05)& {\bf 4.80 (0.96)}& NA\\
\hline
\end{tabular}
}
\end{center}
\end{table}

\subsection{Multi-Mode Simulation}
We also conduct a simulation study where the observed tensor has more than 3 modes.
The setting is similar to Setting 2 in Section \ref{sims}, but with 2 additional modes (i.e., a 5-way tensor) with $d_3=d_4=10$ dimensions.
We use the same parameters as in Setting 2 and adjust $\sigma_e^2$ so that the SNR remains the same.
The results from 100 simulation runs are shown in Table \ref{tab:multi}.
SupCP still provides the best low-rank structure estimation.
We note that the tensor methods (SupCP and CP) tend to have larger variabilities in this multi-mode setting compared to Setting 2.
This may be due to the complexity of multi-mode tensors such as the identifiability issue and the local optimum issue.

We remark that the number of entries in a multiway tensor increases exponentially with the number of modes, presenting a huge challenge in computation.
Existing methods do not scale well to a very large number of modes (say, $>10$).
We deem it a future research direction to extend SupCP to higher modes.

\begin{table}[!h]
\caption{Simulation results for 5-way tensor under Settings 2 (with 100 simulation runs). The median and median absolute deviation (MAD) of each criterion for each method are shown in the table. The best results are highlighted in bold.}
\label{tab:multi}
\begin{center}
{\footnotesize
\begin{tabular}{|c||c|c|c|}
\hline
 Criterion &  SupCP & CP & SupSVD \\
\hline
$SE$    &    {\bf 38.57 (20.79)} & 45.53 (13.80) & 45.05 (0.09)\\
$\angle(\V_1,\widehat{\V_1})$ & {\bf 61.89 (23.23)} & {71.56 (14.11)} & \\
$\angle(\V_2,\widehat{\V_2})$ & {\bf 61.46 (20.29)} & {66.20 (14.90)} &\\
$\angle(\V_3,\widehat{\V_3})$ & {\bf 60.32 (21.56)} & {71.00 (12.45)} &\\
$\angle(\V_4,\widehat{\V_4})$ & {69.04 (14.64)} & {\bf 68.18 (15.41)} &\\
$\|\B-\widehat{\B}\|_F$ & {68.67 (37.90)} & & {\bf 33.95 (0.41)}\\
$100*RE_e$ & {\bf 3.65 (3.45)} & & 5.03 (0.09)\\
$100*RE_f$ & {\bf 23.44 (11.81)}&  & {27.82 (0.92)}\\
Time & 9.67 (3.61) & {\bf 0.63 (0.15)}& 11.28 (0.08)\\
\hline
\end{tabular}
}
\end{center}
\end{table}

\subsection{Misspecified simulation}
\label{misspecified}
Here we describe a simulation study in which the generated data do not have multiway structure.  The setting is analogous to Setting 2 in Section \ref{sims}, except that loadings $\V: 100 \times 5$ are given by the left singular vectors of a $100 \times 5$ matrix of independent $N(0,1)$ entries.   The underlying structure  is then given by the entries of $\U \V^T$, arranged into an array of dimension $100 \times 10 \times 10$.  Thus, $\XX$ has rank $R$ structure when it is matricized, but does not have low rank multilinear structure in $3$ dimensions; i.e., the generative model is supported by the SupSVD model, but not SupCP.  We conduct $100$ replications and estimate a rank $5$ SupCP, CP, and SupSVD fit for each replication. The results are shown in Table~\ref{tab:matrank}, where the principal angle $\angle(\V,\widehat{\V})$ is found for the matricized loadings $\widehat{\V}=\Vmat$ under the SupCP and CP approaches.  As expected, the SupSVD model is superior in this setting, demonstrating that a multiway factorization performs poorly if the data have no multiway structure.

\begin{table}[!h]
\caption{Simulation results for reduced matrix rank with no multiway structure under Settings 2 (with 100 simulation runs). The median and median absolute deviation (MAD) of each criterion for each method are shown in the table. The best results are highlighted in bold.}
\label{tab:matrank}
\begin{center}
{\footnotesize
\begin{tabular}{|c||c|c|c|}
\hline
 Criterion &  SupCP & CP & SupSVD \\
\hline
$SE$    &     93.17 (3.31) & 98.03 (3.02) & {\bf 60.72 (1.71)}\\
$\angle(\V,\widehat{\V})$ &  88.40 (1.17) & 88.84 (1.05) & {\bf 50.25 (9.69)}\\
$\|\B-\widehat{\B}\|_F$ & {137.5 (23.5)} & & {\bf 56.6 (9.96)}\\
$100*RE_e$ &  16.03 (2.54) & & {\bf 6.14 (1.14)}\\
$100*RE_f$ & 67.86 (11.15)&  & {\bf 27.23 (8.87)}\\
Time & 0.469 (0.143) & {\bf 0.095 (0.026)}& 0.213 (0.067)\\
\hline
\end{tabular}
}
\end{center}
\end{table}

\section*{Acknowledgements}

This work was supported in part by National Institutes of Health grant ULI RR033183/KL2 RR0333182 (to EFL).

\bibliographystyle{imsart-number}
\bibliography{suptensor.bib}

\end{document}